# ANALYZES GEOMETRY OF PART AND CHIPS AT THE TIME OF REGENERATIVES VIBRATIONS

Claudiu-Florinel BISU, Jean-Yves K'NEVEZ, Alain GERARD, Raynald LAHEURTE

*Abstract:* *The actions of cut applied to the elastic system cause relative displacements tool/part, which induce a rise in the temperature in the components of the machine tools, its environment, the system-tools-part and generate vibrations. The experimental procedures installation, at the dynamic level, made it possible to determine the elements necessary to a rigorous analysis of the influence of the geometry of the tool, its displacement and evolution of the contacts tool/part and tool/chip on surface carried out*

*Key words: regenerative vibrations, chip variation, self-excited frequencies.*

## 1. INTRODUCTION (10 PT, CAPITALS, BOLD)

The dynamic phenomena of the machine tools come from the interaction of the elastic system machine-process of cut. This interaction is thus the generating source of the dynamic aspects classically met in the machine tools. The effect has been established important ace one off the most sources off sustained relative vibrations between the tool and the workpiece and has been extensively studied in the literature **[1],[2],[4]**.

The actions of cut applied to the elastic system cause relative displacements tool/part, which induce a rise in the temperature in the components of the machine tools, its environment, the system-tools-part and generate vibrations. Those modify the section of the chip, the contact pressure, the relative speed of displacements, etc. Consequently, the instability of the process of cut can cause the instability of the dynamic system of the machine tool: the vibrations are maintained. They have reflected undesirable on the quality of the machined surfaces, their dimensioning, the wear of tool etc. They can generate problems of maintenance even ruptures of elements of the machine tools. Also, it is necessary to develop models making it possible to study the vibratory phenomena met during machining and to envisage the stable conditions of cut.

The modifications of the parameters of the machining system lead to variations of relative displacements tool/part and tool/chip. Those generate variations of the section of the chip which, in their turn, induce variations of the actions of cut, and thus maintain the vibrations. In this part, we present an experimental study to determine and characterize a series of parameters necessary to the dynamic modeling of the cut at the time of the vibrations. In this context, a protocol was conceived using a series of means of three-dimensional measurements to highlight and to measure the various dynamic phenomena, which emerge at the time of the vibratory cut.

## 2. EXPERIMENTAL PROTOCOL

The protocol of tests selected makes it possible to determine a series of phenomenological parameters of the cut. In the case of the three-dimensional cut, the torque of the mechanical actions (forces and, is often truncated because the part couples (moment) torque is generally neglected fault of metrology adapted to their evaluation. However, an experimental study undertaken by Cahuc et al. [5] showed that the taking into account of the moments of cut allows a better evaluation of the consumption and can, a priori, to make it possible to act on the wear of the cutting tools. The use of a dynamometer measuring the six components torque of the mechanical actions makes it possible to evaluate the dynamic influence of the vibrations of cut on system POM (part/tool/machine). Like the behavior of the system machining present couplings in all the directions, the quantification of the dynamic movement of the tool during the cut is carried out in a three-dimensional way, thanks to an accelerometer 3D.

### 2.1 Experimental device

The tests of dynamic cut are carried out on a conventional lathe (fig.1). The behavior is identified with a three-dimensional accelerometer fixed on the tool and two unidimensional accelerometers positioned on the lathe, side stitches, to identify the influence of the pin on the process. The efforts and the moments (couples) with the point of the tool are measured using a dynamometer with six components, while the evolution the instantaneous speed of the part is given by a rotary coder. The three-dimensional dynamic character is highlighted by seeking the various existing correlations or the various evolutions of the parameters, which make it possible to characterize the process.

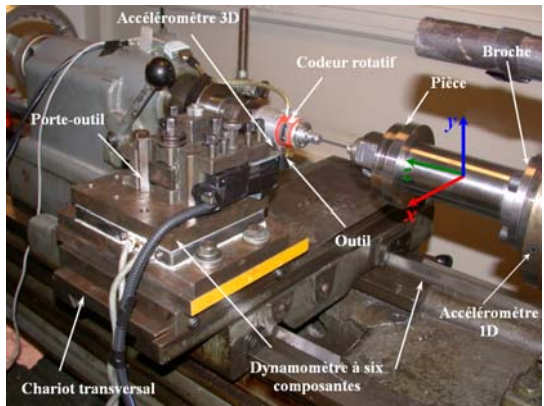

**Fig.1**. Experimental device

This analysis provides the necessary information to the design and the development of model of the three-dimensional dynamic cut. Thus, the experimental study of the phenomena met is essential to obtain a model reflecting the reality of the processes present at the time of the cut.

**2.2 Protocole d'essais**

The method of characterization of the self-sustained vibrations is based on various work [1], [2], [3]. In order to determine the zone of the self-sustained vibrations, the parameters of cut are selected constant except the depth of cut. The self-sustained vibrations are almost non-existent for ap=1mm. Their appearance is clearly observed for ap=5mm. Around this point of operation, of many tests are carried out, to determine the extent of the self-exited vibrations. These tests are carried out by maintaining the same value of ap and same number of revolutions *(N)* of the part and for various advances *(f)*, table 1. On the testing ground, the instantaneous speed of rotation is controlled by a rotary coder directly related to the part. The connection is carried out by a rigid steel wire, which makes it possible to have a better transmission of the behavior (fig.1). We observe during the cut that the number of revolutions remains constant and is located at approximately of 690 tr/min. We detect a variation cutting speed of about 1%, which can be neglected. For these tests, the tool used is of type TNMA 160412 out of carbides not covered without breeze chip and the machined material is of 42CrMo4 type.

*Table 1*
**Cutting conditions.**

| ap (mm) | N (tr/min) | f (mm/tr) | | | |
|---|---|---|---|---|---|
| 5 | 690 | 0.05 | 0.0625 | 0.075 | 0.1 |

*Table 2*
**Geometrical characteristics of the tool.**

| $\gamma$ | $\alpha$ | $\lambda_s$ | $\kappa_r$ | $r_\varepsilon$ | R |
|---|---|---|---|---|---|
| -6° | 6° | -6° | 91° | 1,2 mm | 0,02 mm |

The test-tubes are of a diameter of 120 mm and a 30 mm length (Fig.2). Dimensions of the door part were optimized by method finite elements in order to ensure maximum rigidity compatible with self excited frequencies of the crew sufficiently isolated of the frequencies of vibrations induced by has cut.

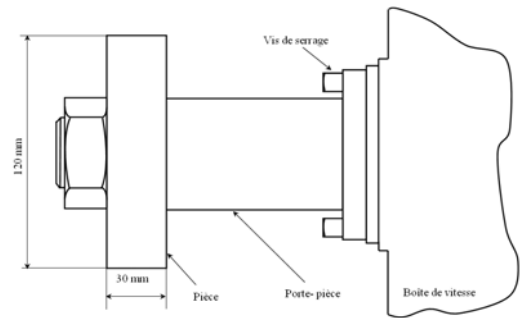

**Fig.2**. Test part

The geometry of the tool (plate) is characterized by the angle of cut *(γ)*, the clearance angle *(α)*, the angle of inclination of edge *($\lambda_S$)*, the angle of attack, the radius of nose *($r_\varepsilon$)* and the radius of acuity *(R)*. In order to limit to the maximum the appearance of wear along the face of cut, the plate is examined after each test and is changed if necessary.

**3. INTERACTION OUTIL/PIECE**

Part BO is regarded as an interdependent block, which is subjected to the vibrations due to the process of cut. The machining system represents a closed system; interaction BO and BP carries out the closing of this system. The appearance of the vibrations is strongly conditioned by the behavior of the elastic structure of the system. We present an analysis of the measured sizes. This one is divided into two great parts:
• the first consists of measurements of accelerations and evolution of displacements;
• the second refers to the measurement of the mechanical actions, the complete torque. Lastly, we report the results relating to the process of cut, measurements of roughness on the part and the geometrical characterization of the chips.

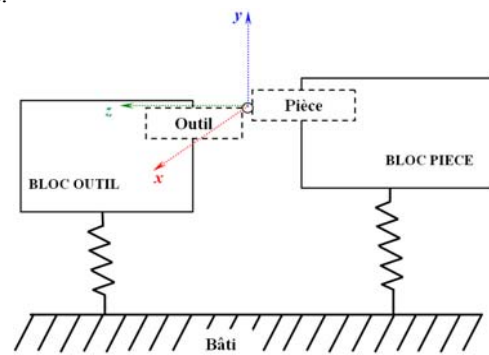

**Fig.3**. **BO** – **BP** interaction.

The tests out of three-dimensional cut are carried out under the conditions presented to the preceding paragraph. The vibrations at the time of the cut are clearly observed for a depth of cut ap=5mm (fig. 4). Tests are also carried out for a depth of cut of ap=1mm, 2mm or 3mm, in order to compare the signals and to dissociate the influence of the vibrations on the cut itself.

During these tests for ap=1mm, 2mm or 3mm the system is stable, the signal has very small amplitudes about the micrometer, and the section of the chip remains constant. The surface quality of the part is correct, with a total roughness *(Rt)* of 4.3 μm. The increase depth of cut up to 5 mm, makes it possible to reach a mode of unstable cut. We also make vary the advances around this value. In this situation, the chips present periodic undulations and variations of the section seem a discontinuous cut in the case of [3]. The characterizations of the chip and the part are carried out in the following paragraph.

In order to analyze displacements in the three directions, we determine via frequential analyses by FFT (Fast Fourier Transformation), the frequential spectra on the three components of accelerations. We eliminate the frequencies with the tops of 1000 Hz, knowing that in the literature the beach of frequency of the self-sustained vibrations lies between 120 Hz and 200 Hz, [3], [4], [5]. The measurement of the vibrations during the machining led to the results presented to the fig. 4. The frequencies of the vibrations of cut are around 190 Hz for the three axes with a dominating amplitude on the axis $y$ (axis of cut). We consider during the analysis that the recorded frequencies correspond well to the regenerative vibrations related to the phenomenon of the cut. The peack of frequency around 200 Hz is found for the other tests according to the advance.

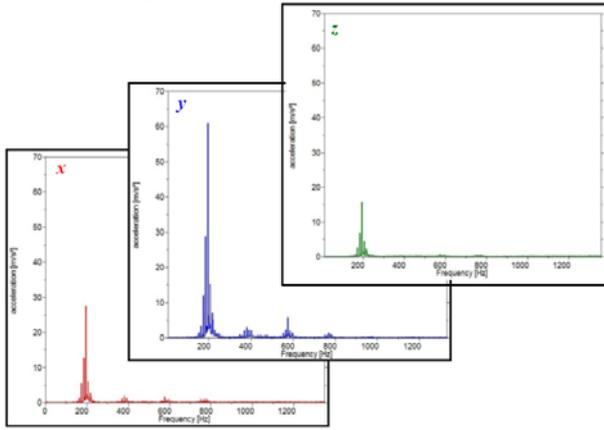

**Fig.4** : FFT of the signals of accelerations according to the three directions.

The regenerative vibrations have an influence on the surface quality of the parts (fig.5). Analysis FFT of the rugosimetric data shows a peak of frequency located around 190 Hz, which is coherent with the preceding data.

## 4. PART AND CHIP GEOMETRY

### 4.1 Rugosimetric measurements

The regenerative vibrations have an influence on the surface quality of the parts (fig.5). Analysis FFT of the rugosimetric data shows a peak of frequency located around 200 Hz, which is coherent with the preceding data.

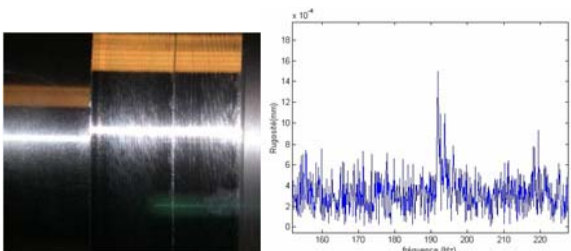

**Fig.5**. FFT of the profile of roughness of the machined part.

### 4.2 Characteristics of the chip

Microscopic measurements were carried out on the chips and made it possible to determine the variations thickness and width of those. Variations between the maximum thickness and the minimal thickness are about 2, and independent of the advance. An example is presented (fig.6) for a sample of chip during a test with an advance of 0.05 mm/tr where $h$ =0.23mm and $h$= 0.12mm.

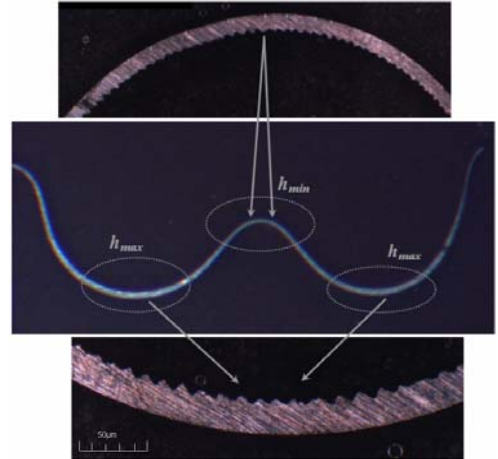

**Fig.6**. Variation thickness of the chip.

The measure of length of the chip corresponding to an undulation enables us to find the frequencies of self-sustained vibrations starting from the cutting speed. To determine the overall length of the chip, it is necessary to measure the wavelength and to take account of the rate of the chip at the time of the cut.

$$f_{cop} = \frac{V}{l_o \cdot \xi_c}, (1)$$

with $f_{cop}$ the frequency of the chip, $V$ the speed of the chip, $l_o$ the length of an undulation of chip and $\xi_c$ the coefficient of work hardening for the chip. The coefficient of work hardening is given using the it is then validated by the literature.

$$\xi_c = \frac{\cos(\phi-\gamma)}{\sin(\phi)}, (2)$$

In our case, a 11 mm length of undulation is measured on the chip, with a coefficient of work hardening $(\xi_c)$ from 1.8 and one cutting speed $(V)$ of 238 m/min. We obtain then, a frequency of 206 Hz, very near to the frequencies of displacements or forces raised at the time of the cut. The width of the chip is then measured with the same techniques. Important variations, about 0,5mm, are observed. The maximum width $(w_{max})$ is of 5.4 mm and the minimal width $(w_{min})$ 4.9 mm, (fig.7). The measurement of the angle (fig.7) of the slopes of the width of the chip between each undulation, close to 29°, is equal to the dephasing measured on the signals of displacements (fig.8).

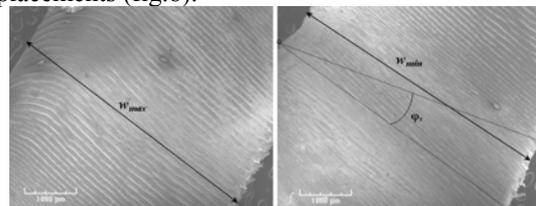

**Fig.7**. Variation of the width of the chip.

## 5. SELF-EXCITED VIBRATION, EXPERIMENTAL VALIDATION

According to this study we consider that the zone of self-sustained vibrations is around 190 Hz. The analysis carried out to the measures of displacements and the forces makes it possible to evaluate existing dephasing between forces and displacements. This dephasing explains the delay of the force compared to displacement. The appearance of regenerative vibrations can be also explained by the delay forces/displacement, which increases the energy level in the system. The existence of this delay is explained by the inertia of the machining system and more particularly by the inertia of the process of cut [1].

*Table 3*
**Valeurs du déphasage force / déplacement.**

| $\varphi_{fu\_x}$ | $\varphi_{fu\_y}$ | $\varphi_{fu\_z}$ |
|---|---|---|
| 13° | 23° | 75° |

Dephasing forces/displacement remains constant according to the advance. The appearance of the self-sustained vibrations thus remains explicable by dephasing between the forces and displacements. When the tool moves on the part $e_1,e_2,e_3$ ellipse, the cutting force carries out a positive work because its direction coincides with the direction of cut. On the part $e_3,e_4,e_1$, the work produced by the cutting force is negative. By comparing the two parts of the ellipse, we notice that the force on the course $e_1,e_2,e_3$ is larger than on the course $e_3,e_4,e_1$ because the depth of cut is larger. During this process, work remains positive and the resulting energy stored by the system allows the maintenance of the vibrations.

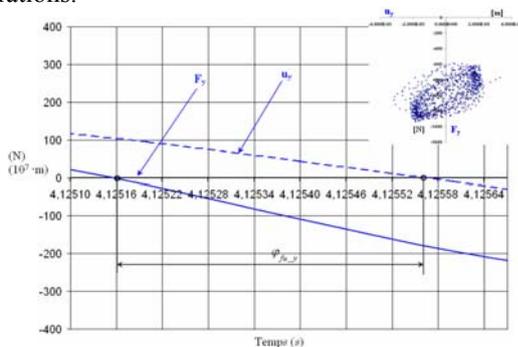

**Fig.8**. Dephasing forces/displacements according to Y.

## CONCLUSIONS

The experimental procedures installation, at the dynamic level, made it possible to determine the elements necessary to a rigorous analysis of the influence of the geometry of the tool, its displacement and evolution of the contacts tool/part and tool/chip on surface carried out. In addition, the coupling highlighted between the elastic characteristics of the machining system and the vibrations generated by the cut made it possible to establish that the appearance of the self-sustained vibrations is strongly influenced by the stiffnesses of the system, their report/ratio and their direction. We also established a correlation between the direction of the vibratory movement of the elastic structure of the machine tool and the variations of section of the chip.

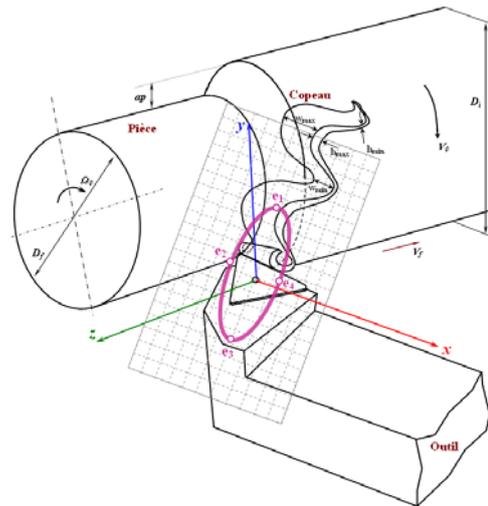

**Fig.11**. Elliptic trajectory of the movement the tool/part

The self-sustained correlations part-chips/vibrations are valid. These first results make it possible to now consider a more complete study by exploiting the concept of torque completely. Indeed, thanks to the dynamometer with six components, we identified the existence of moments to the point of the tool which we did not present here and who are not evaluated by the traditional measuring equipment. In the long term the originality of work partly presented here relates to the analysis of the torque of complete actions applied to the point of the tool, with an aim of making evolve/move a model semi analytical 3D of the cut.

## REFERENCES

[1] J. Tlusty, *Analysis of the state of research in cutting dynamics*, Ann. CIRP 27 (1978) 583–589.
[2] G. Stépán, *Modelling non-linear regenerative effects in metal cutting,* Proc. R. Soc. London A 359 (2001) 739–757. Technical Conference, Sacramento, CA, USA, 1997.
[3] C.F. Bisu (2007), *Etudes des vibrations auto-entretenues en coupe trisimensionnelle: nouvelle modélisation appliquée au tournage*, Thèse de doctorat, Bordeaux.
[5] P. Wahi, A. Chatterjee, *Self-interrupted regenerative metal cutting in turning,* Int. J. of Non-Lin. Mech.43 (2008) 111 – 123
[1] Cahuc O., Darnis P., Gerard A., Battaglia J., (2001) , *Experimental and analytical balance sheet in turning applications,* Int. Journal of Advanced Manufacturing Technologies, Vol 18, No.9, pp. 648-656.

**Author(s):**

Claudiu-Florinel BISU, University POLITEHNICA of Bucharest, Machines ans Systems of Production Departement, cfbisu@gmail.com
Alain GERARD, University of Bordeaux, Mechanics and Physics Laboratory , alain.gerard@u-bordeaux.fr.
Jean-Yves K'NEVEZ, University of Bordeaux, Mechanics and Physics Laboratory jean-yves.knevez@u-bordeaux.fr
Raynald LAHEURTE, University of Bordeaux, Mechanics and Physics Laboratory , raynald.laheurte@u-bordeaux.fr.